\documentclass{aa}
\usepackage{graphicx}
\input psfig
\begin{document}
   \title{The low-level radial velocity 
          variability in \\
          Barnard's star (=GJ~699) 
   \thanks{Based on observations collected at the 
           European Southern Observatory, Paranal, Chile (ESO programmes
           65.L-0428, 66.C-0446, 267.C-5700, 68.C-0415,
           and 69.C-0722)
}
          }

   \subtitle{Secular acceleration, indications for convective redshift,\\
    and planet mass limits}

   \author{M. K\"urster \inst{1} \and
           M. Endl \inst{2} \and
           F. Rouesnel\inst{3} \and
           S. Els \inst{4} \and
           A. Kaufer \inst{5} \and \\
           S. Brillant \inst{5} \and
           A.P. Hatzes \inst{1} \and
           S.H. Saar \inst{6} \and
           W.D. Cochran \inst{2}
          }

   \offprints{M.~K\"urster:}

   \institute{Th\"uringer Landessternwarte Tautenburg,
              Sternwarte 5, D-07778 Tautenburg, Germany\\
              \email{M.K\"urster: martin@tls-tautenburg.de,
                     A.P.Hatzes: artie@tls-tautenburg.de}
         \and
              McDonald Observatory, The University of Texas at Austin,
              Austin, TX 78712-1083, USA\\
              \email{M.Endl: mike@astro.as.utexas.edu,
                     W.D.Cochran: wdc@shiraz.as.utexas.edu
                     }
        \and 
              Universit\'e de Paris-Sud, B\^atiment 470,
              F-91405 Orsay Cedex, France\\
              \email{F.Rouesnel: frederic.rouesnel@obspm.fr}
        \and
              The Isaac Newton Group of Telescopes, Apartado 321,
              E-38700 Santa Cruz de La Palma, Canary Islands, Spain\\
              \email{S.Els: sels@ing.iac.es}
        \and
              European Southern Observatory, Casilla 19001, Vitacura,
              Santiago 19, Chile\\
              \email{A.Kaufer: akaufer@eso.org,
                     S.Brillant: sbrillan@eso.org}
        \and
              Harvard-Smithsonian Center for Astrophysics, 60 Garden
              Street, Cambridge, MA 02138, USA\\
              \email{S.H.Saar: ssaar@cfa.harvard.edu}
             }

   \date{Received 10 February 2003 / Accepted 11 March 2003}

   \abstract{
We report results from $2~1/2~{\rm yr}$ of high precision
radial velocity (RV) monitoring of Barnard's star. 
The high RV measurement precision of the VLT-UT2+UVES
of $2.65~{\rm m~s}^{-1}$ made the following findings possible.
(1)~The first detection of the change in the RV of a star caused
by its space motion (RV secular acceleration).
(2)~An anti-correlation of the measured RV with the strength of the
filling-in of the ${\rm H}_\alpha $ line by emission.
(3)~Very stringent mass upper limits to planetary companions.
Using only data from the first 2 years, we obtain a best-fit value for the
RV secular acceleration of 
$5.15\pm 0.89~{\rm m~s}^{-1}~{\rm yr}^{-1}$. This agrees within $0.95\sigma $
with the predicted value of $4.50~{\rm m~s}^{-1}~{\rm yr}^{-1}$
based on the Hipparcos proper motion and parallax combined
with the known absolute radial velocity of the star.
When the RV data of the last half-year are added the best-fit slope
is strongly reduced to
$2.97\pm 0.51~{\rm m~s}^{-1}~{\rm yr}^{-1}$ ($3.0\sigma $ away from
the predicted value), clearly suggesting the presence of
additional RV variability in the star.
Part of it can be attributed to stellar activity as we demonstrate by
correlating the residual RVs with an index that describes
the filling-in of the H$_\alpha $ line by emission. A
correlation coefficient of $-0.50$ indicates that the appearance of
active regions causes a blueshift of photospheric absorption lines.
Assuming that active regions basically inhibit convection we discuss
the possibility that the fundamental (inactive) convection pattern
in this M4V star produces a convective redshift which would indicate that
the majority of the absorption lines relevant for our RV measurements
is formed in a region of convective 
overshoot. This interpretation could possibly extend a trend
indicated in the behaviour of earlier spectral types that exhibit convective
blueshift, but with decreasing line asymmetries and blueshifts as one
goes from G to K dwarfs. Based on this assumption, we
estimate that the variation of the
visible plage coverage is about $20\% $. We also determine
upper limits to the projected mass $m\sin i$ and to the true mass $m$
of hypothetical planetary companions in circular orbits.
For the separation range $0.017-0.98~{\rm AU}$ 
we exclude any planet with 
$m\sin i> 0.12~{\rm M}_{\rm Jupiter}$ and $m> 0.86~{\rm M}_{\rm Jupiter}$.
Throughout the habitable zone around Barnard's star,
i.e.~$0.034-0.082~{\rm AU}$, we exclude planets with 
$m\sin i> 7.5~{\rm M}_{\rm Earth}$ and $m> 3.1~{\rm M}_{\rm Neptune}$.

       \keywords{techniques: radial velocities --- stars: kinematics
	 --- stars: planetary systems --- stars: activity}
   }

   \titlerunning{The low-level radial velocity variability in Barnard's star}
   \authorrunning{M. K\"urster et al.}
   \maketitle
%

%------------------------------------------------------------------------------
\section{Introduction}
%------------------------------------------------------------------------------

The search for extrasolar planets via precise stellar radial velocity (RV)
measurements has led to more than 100 discovery announcements up to now
(e.g.~Mayor \& Queloz 1995; Marcy \& Butler 1996; K\"urster et al.~2000).
A good fraction of these discoveries is summarized in the paper by
Butler et al.~(2002). The majority of the extrasolar planets has so
far been found around main sequence stars of spectral types from
late--F through K. There is only one M dwarf with planetary companions
known, GJ876, which is orbited by two Jovian-type planets with
orbital periods of $30.1$ and $61.0~{\rm day}$, respectively 
(Marcy et~al.~2001; Delfosse et al.~1998). 
Combining RV data with new astrometric
measurements Benedict et~al.~(2002) determine a mass of 
$1.89\pm 0.34~{\rm M}_{\rm Jupiter}$ for the outer planet.

It is due to the intrinsic faintness of M dwarfs that this most
abundant type of star tends to be underrepresented in RV searches
as very high measurement precision requires large telescopes. This
introduces a strong bias in any attempt to estimate the frequency of
planetary companions to stars as a whole.

On the other hand, due to the low mass of M dwarfs an orbiting planet of
a given mass causes a higher reflex motion of the star. Therefore, the
present day measurement precision for differential radial velocities
(DRV)
\footnote{Throughout this paper we will differentiate between
{\em absolute} radial velocity (ARV) and {\em differential} radial
velocity (DRV), the latter being relative to any one of the
measurements. It is the DRV for which high precision can be achieved
and only DRVs are needed when RV changes are to be studied.}
of a few
${\rm m~s}^{-1}$ holds the prospect to discover terrestrial planets
(with a few Earth masses) in close-in orbits around the lower mass M dwarfs,
e.g.~in their habitable zones (Schneider 1998; Endl et~al.~2003).

For instance, for a main sequence star with a 
mass of $0.16~{\rm M}_\odot $, such as
that of Barnard's star, the subject of the present paper (mass
from Henry et~al.~1999), the habitable zone lies at 
$0.034-0.082~{\rm  AU}$ (see Kasting et al.~1991) corresponding to
orbital periods in the range $5.75-21.5~{\rm d}$.
Around a star of this mass a planet in a circular orbit with a
projected mass $m\sin i=1~{\rm M}_{\rm Earth}$ would cause a
peak-to-peak RV amplitude of $2.4~{\rm m~s}^{-1}$ at the inner edge
of the habitable zone and $1.6~{\rm m~s}^{-1}$ at its outer edge.
For $m\sin i=3~{\rm M}_{\rm Earth}$ the resulting
peak-to-peak RV amplitudes would be
$7.3~{\rm m~s}^{-1}$ and $4.7~{\rm m~s}^{-1}$, respectively.

When attempting to measure such small RV signals it becomes important
to consider other effects that affect the
determination of DRVs. Two such effects are
important for M dwarfs. First, the high proper motion of
nearby stars causes a secular change in their RV which for some
stars can reach several ${\rm m~s}^{-1}$ within a few years.
Second, intrinsic variability due to stellar activity can mimic
RV variations. Little is known so far about the
level at which this complicates DRV measurements in M dwarfs.

This paper reports on high precision DRV measurements obtained with 
VLT-UT2+UVES of Barnard's star which, as we demonstrate, 
exhibits a considerable secular change of its RV.
It is known as a quite inactive star and was selected with the
expectation that its DRV measurements are largely unaffected by
activity; as we shall see this is not the case.

The paper is structured as follows. We summarize the known properties
of Barnard's star (Sect.~2) and proceed with a description of the concept
of RV secular acceleration (Sect.~3). Our observations are described in
Sect.~4 along with a brief outline of our data modelling technique.
In Sect.~5 we present results on the RV secular acceleration,
on the relation between DRVs and stellar activity as evidenced in the
strength of the 
H$_\alpha $ line, and on period analysis in the RV and 
${\rm H}_\alpha $ data;
we also provide mass upper limits to planetary companions.
We discuss our results in Sect.~6 and summarize our conclusions in
Sect.~7.

%--------------------------------------------------------------------------
   \begin{figure}
%   \sidecaption
   \resizebox{\hsize}{!}{\includegraphics{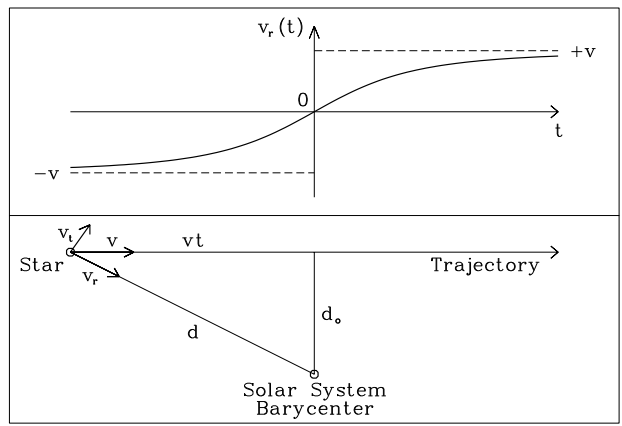}}
   \centering
      \caption{Lower panel: Geometry of the space motion of a passing star.
               $v_{\rm t}$ and $v_{\rm r}$ are, respectively,
               the transversal and
               radial components of the space velocity $v$.
               The current distance is $d$, the minimum
               distance is $d_\circ $, and $vt$ is the
               trajectory from the current position to the
               point of nearest approach.
               Upper panel: The ARV as a function of time
               and its asymptotic limits $\pm v$.
%%% $\lim_{t \rightarrow \pm \infty} v_{\rm rad}=\pm v$.
              }
%         \label{Fig1}
   \end{figure}
%--------------------------------------------------------------------------

%--------------------------------------------------------------
\section{Properties of Barnard's star}
%------------------------------------------------------------------------------

The M4V dwarf Barnard's star (GJ~699, HIP 87937) 
is the nearest star after Proxima Centauri and the $\alpha $~Centauri
binary. Its parallax has been determined by Hipparcos as 
$549.01\pm 1.58~{\rm mas}$ (ESA 1997; Perryman et~al.~1997),
i.e.~the distance is $1.8215\pm 0.0052~{\rm pc}$.
% More recently, Benedict et~al.~(1999), using the HST FGS,
% determined $545.4\pm 0.3~{\rm mas}$ corresponding to $1.834~{\rm pc}$.
Barnard's star
has long been known as the highest proper motion star
in the sky (Barnard 1916, 1917). For epoch 1991.25 Hipparcos gives
$\mu _\alpha =-797.84~{\rm mas~yr}^{-1}$
and $\mu _\delta =10326.93~{\rm mas~yr}^{-1}$ for the proper motion in
right ascension and declination, respectively (ESA 1997). The tangential
component of the space velocity (perpendicular to the line of sight)
is thus $v_{\rm t}=89.440~{\rm km~s}^{-1}$. The secular change
of the proper motion is $1.2~{\rm mas~yr}^{-2}$
(Gatewood 1995).
%%% (Gatewood 1995; Benedict et al.~1999).

Barnard's star also has a high ARV.
Dravins et~al. (1999) find an astrometrically derived ARV
of $v_{\rm r}=-101.9\pm 6.5~{\rm km~s}^{-1}$, somewhat smaller 
(in absolute value) than the spectroscopic
value of $-110.506~{\rm km~s}^{-1}$ by Nidever et~al.~(2001) 
who estimate their systematic errors 
in their treatment
of gravitational redshift and convective blueshift) to
$0.4~{\rm km~s}^{-1}$.
Other spectroscopic literature values
are, e.g., $-110.9\pm 0.2~{\rm km~s}^{-1}$
(Garc\'\i a-Sanchez et al. 2001) and
$-109.7~{\rm km~s}^{-1}$ (Turon et al.~1998).
\footnote{The Hipparcos and Tycho Catalogues give the ARV value used
to correct for the astrometric perspective acceleration
(change in proper motion) of Barnard's star as $-111.0~{\rm km~s}^{-1}$.}
Thus the overall space velocity is
$v\approx 142~{\rm km~s}^{-1}$.
In the following we will adopt the ARV value by Nidever et~al.~(2001)
noting that none of the results presented in this paper
would be significantly altered, if, e.g., we
took the astrometric ARV value by Dravins et~al.~(1999).
The same would be true, if we replaced the Hipparcos astrometric data
by the recently refined values by Benedict et~al.~(1999) who used the
% the HST FGS
Fine Guidance Sensor~3 of the Hubble Space Telescope
for interferometric fringe tracking astrometry.

Due to its high space motion and sub-solar metallicity Gizis (1997)
classifies Barnard's star as an ``Intermediate Pop.~II star''
(following Majewski 1993), i.e.~an
object between a halo star and a thin disk star. 
Its low observed X-ray luminosity (Marino et~al.~2000; H\"unsch et~al.~1999)
and its presumable rotation period of $130.4~{\rm d}$
(from weak evidence for $\approx 0\fm 01$ 
photometric variations by Benedict et~al.~1998) also indicate an inactive
old star.

%---------------------------------------------------------------
   \begin{table}
   \centering
      \caption{Kinematic data for Barnard's star. For all data
               used in our calculation (Hip = Hipparcos,
               Nid = Nidever et~al.~2001, UV = our UVES data)
               we list the epoch,
               the distance $d$, the ARV $v_{\rm r}$, and the secular
               acceleration of the RV ${\rm d}v_{\rm r}/{\rm d}t$.
               We also compare our results for the point of closest
               approach (CA) with those of Garc\'\i a-Sanchez
               et~al.~2001 (GS).
}
      \vspace{0.2cm}
         \begin{tabular}{lllcc}
            \hline
            \noalign{\smallskip}
Data/ & Epoch & $~~~~d$ & $v_{\rm r}$ & ${\rm d}v_{\rm r}/{\rm d}t$ \\
Event
& & $~~[{\rm pc }]$ & $[{\rm kms}^{-1}]$ & $[{\rm ms}^{-1}{\rm yr}^{-1}]$\\
            \noalign{\smallskip}
            \hline
            \noalign{\smallskip}
Hip     & $~1991.25$ & $1.8215$ & $-110.532$ & $~4.491$ \\
Nid     & $~1996.91$ & $1.8209$ & $-110.506$ & $~4.496$ \\
UV      & $~2001.51$ & $1.8203$ & $-110.485$ & $~4.499$ \\
            \noalign{\smallskip}
            \hline
            \noalign{\smallskip}
CA      & $11730$ & $1.1458$ & $~~~0.0$ & $18.043$ \\
GS      & $11700_{^{\pm 100}}$ & $1.144_{^{\pm 0.005}}$ & $~~~(0.0)$
          & not given\\
%GS2002 & $11700^{\pm 100}$ & $1.144^{\rm \pm 0.005}$ & $~~~0.0$ &  not given\\
%GS2001   & $11700~~~~$ & $1.144~$ & $~~~0.0$ &  not given \\ 
%          & $\pm 100$ & $\pm 0.005$ & &  \\ 
             \noalign{\smallskip}
           \hline
         \end{tabular}
   \end{table}
%---------------------------------------------------------------

For many years Barnard's star was also a prime candidate for displaying 
astrometric pertubations attributable to one or two planetary companions in
long-period orbits of about $12$ and $20~{\rm yr}$
% $\approx 12~{\rm yr}$ and $\approx 20~{\rm yr}$
(van de Kamp 1963, 1982). This was later refuted by Gatewood (1995).
The most recent astrometric measurements by Benedict et al.~(1999)
placed stringent mass upper limits of $2.1~{\rm M}_{\rm Jupiter}$ down to
$0.37~{\rm M}_{\rm Jupiter}$ for orbital periods of $50~{\rm d}$ up to
$600~{\rm d}$ (see Fig.~8 in Sect.~5.4).

%------------------------------------------------------------------------------
\section{The secular acceleration of the radial velocity}
%------------------------------------------------------------------------------

The proximity and high space velocity of Barnard's star implies
a detectable secular change of its radial velocity (RV secular
acceleration; van de Kamp 1977).
To illustrate this effect
Fig.~1 (lower panel) shows the trajectory past the barycenter of the
solar system of a star in linear space motion,
i.e.~ignoring the subtle variations of
the galactic potential (see Garc\'\i a-Sanchez et~al.~2001).
The minimum distance $d_\circ $ at the point of closest approach
is given by
\begin{equation}
%%%   d_\circ = d / \sqrt{1 + v_{\rm r}^2 / v_{\rm t}^2}~,
   d_\circ = \frac{d}{\sqrt{1 + v_{\rm r}^2 / v_{\rm t}^2}}~,
\end{equation}
where $d$, $v_{\rm t}$, and $v_{\rm r}$ are the distance,
and the tangential and radial velocity components, respectively,
at some earlier epoch. The time $t$ at
the earlier epoch is
\begin{equation}
%%%   t = t_{\circ }-\sqrt{d^2 - d_\circ ^2}/v~,
   t = t_{\circ }-\frac{\sqrt{d^2 - d_\circ ^2}}{v}~,
\end{equation}
where $v=\sqrt{v_{\rm r}^2 + v_{\rm t}^2}$ is the space velocity
and $t_{\circ }$ is the time at closest approach. Setting $t_{\circ }=0$
the ARV as a function of time is given by
\begin{equation}
%%%   v_{\rm r}(t) = v^2 t/\sqrt{v^2 t^2 + d_\circ ^2}~.
   v_{\rm r}(t) = \frac{v^2 t}{\sqrt{v^2 t^2 + d_\circ ^2}}~.
\end{equation}
The upper panel of Fig.~1 shows a graph of this function.
Differentiation yields the secular acceleration of the RV:
%\begin{eqnarray}
%   \frac{{\rm d}v_{\rm r}(t)}{{\rm d}t} & = & 
%   \frac{V^2}{\sqrt{V^2 t^2 + d_\circ ^2}} -
%   \frac{V^4t^2}{(V^2 t^2 + d_\circ ^2 )^{3/2}} \\
%   & = &
%   4.499~{\rm m~s^{-1}~yr^{-1}}.\nonumber
%\end{eqnarray}
\begin{equation}
   \frac{{\rm d}v_{\rm r}(t)}{{\rm d}t} = 
   \frac{v^2}{\sqrt{v^2 t^2 + d_\circ ^2}} -
   \frac{v^4t^2}{(v^2 t^2 + d_\circ ^2 )^{3/2}}~.
\end{equation}
Kinematic data for Barnard's star using Eqs.~(1--4) with the input data from
Sect.~3 are
summarized in Table~1 that lists the distance, the ARV, and its secular
acceleration for the individual epochs of all measurements relevant to
this paper. 
\footnote{To calculate the RV secular acceleration we
combined the Hipparcos proper motion and distance with the ARV value
by Nidever et~al.~(2001) transformed to the same epoch.}
For the time of closest approach
to the solar system our results are compared with those
of Garc\'\i a-Sanchez (2001).

At the mean time of our UVES observations presented here (epoch 2001.51)
the RV secular acceleration is $4.499~{\rm m~s^{-1}~yr^{-1}}$.
Over the time baseline for our UVES data of
$2.422~{\rm yr}$ the acceleration value increases by only 
$0.0020~{\rm m~s^{-1}~yr^{-1}}$;
we neglect this small increase and treat the secular acceleration
as a purely linear function when comparing it to our UVES data.

% --------------------------------------------------
   \begin{figure*}
   \sidecaption
   \includegraphics[width=12cm]{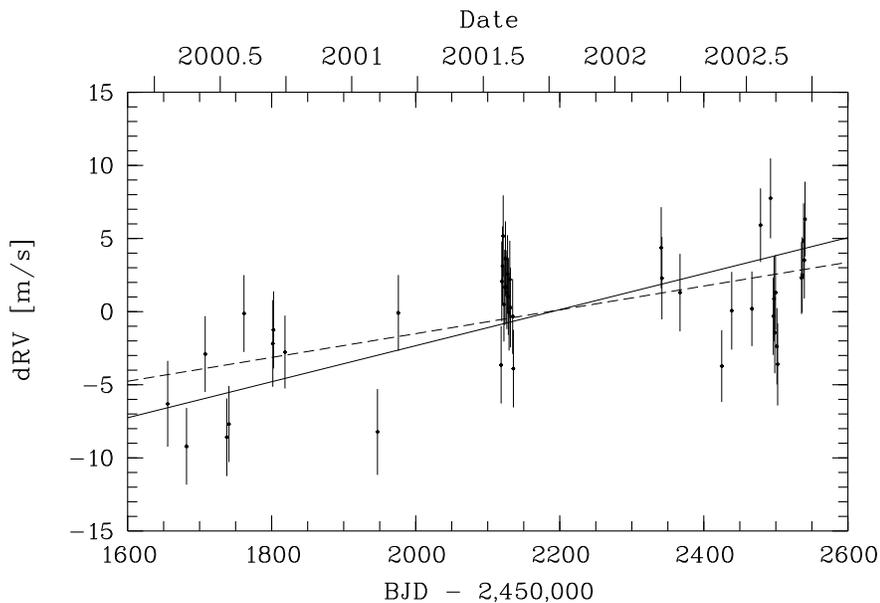}
%
%   \resizebox{\hsize}{!}{\includegraphics{figure2.eps}}
%
%   \centering
%   \includegraphics[width=10.0cm]{figure2.eps}
%%%   \includegraphics{figure2.eps}
      \caption{~Time series of the differential RV data of Barnard's
      star plotted vs.~Barycentric Julian Day (bottom) and the Date (top).
      The solid line represents the predicted RV secular acceleration of
      $4.499~{\rm m~s^{-1}~yr^{-1}}$. The dashed line is a linear fit to
      the data with a
      slope of $2.97\pm 0.51~{\rm m~s}^{-1}~{\rm yr}^{-1}$.
              }
%         \label{Fig2}
   \end{figure*}
%
%--------------------------------------------------------------

%------------------------------------------------------------------------------
\section{Observations and data modelling}
%------------------------------------------------------------------------------

Barnard's star was observed with the VLT-UT2+UVES in the
framework of our ongoing precision RV survey of M dwarfs in search for
extrasolar planets. UVES was used together with its self-calibrating
iodine gas absorption cell operated at a temperature of $70^{\circ }{\rm C}$.
Image slicer \#3 and an $0.3^{\prime \prime }$ slit were chosen
yielding a resolving power of $R=100\,000-120\,000$.
At the selected central wavelength of $600~{\rm nm}$
the useful spectral range containing iodine $(\mathrm{I}_2)$
absorption lines ($\approx 500 - 600~{\rm nm}$) falls 
entirely on the better quality CCD of the mosaic of two
$4~{\rm K}\times 2~{\rm K}$ CCDs.
% This EEV CCD collects the wavelength range $491.9-603.2~{\rm nm}$ in
% our setup, whereas the second CCD, an MIT devive that is known to degrade
% the resolving power and that is also cosmetically inferior,
% collects the wavelength range $599.1-707.1~{\rm nm}$.

Our analysis of the stellar spectra to determine differential radial
velocities (DRV) followed Endl~et~al.~(2000). Briefly, 
DRV measurements were made with the iodine cell
inserted in the light path 
which produces a superposition of a forest of dense iodine
absorption lines onto the stellar spectrum. A higher-resolution
$(R=700\,000-1\,000\,000)$ template spectrum of the iodine cell 
obtained with the McMath Fourier Transform Spectrometer (FTS) on
Kitt Peak was
used to reconstruct the instrumental profile (IP) from
observations of rapidly rotating B-stars through the iodine cell.
This was required in one night for Maximum Entropy deconvolution
(with the IP) of a pure stellar spectrum (taken without the iodine
cell) in order to obtain a stellar template spectrum with a resolution
matching that of the FTS iodine spectrum.
Finally, both template spectra were combined to model
the regular observations (star+iodine), again employing IP reconstruction.
From the Doppler shift of these model spectra with respect to the
stellar template we obtain a time series of the DRV of the star which
is largely corrected for instrumental instabilities.

As the IP varies strongly within an Echelle order of UVES spectra as
well as across the orders, the modelling was done in small spectral
chunks of $\approx 0.2~{\rm nm}$ each of which yields an independent
DRV measurement.
DRV values from $\approx 580$ spectral segments (after rejection of
outliers) were then averaged to produce
the final DRV value for the spectrum; the error of this mean value was
taken as the measurement error.

A total of 137 star+iodine spectra are available from 46 nights between 
21 April 2000 and 23 Sept.~2002. 
Triplets of exposures were obtained each night within about 16~min,
except for one night where only two spectra could be secured.
Individual exposure times were $250-285~{\rm s}$ yielding a typical
S/N per pixel of $66$. On average our error of the individual RV
measurement is $2.65~{\rm m~s}^{-1}$ demonstrating the high precision
achievable with UVES on this faint star $(V=9.54~{\rm mag})$.
All RV data were corrected to the solar system bary\-center
using the JPL ephemeris DE200
(Standish 1990) for the flux-weighted temporal midpoint of the exposure
as given by the UVES exposuremeter. 
For each epoch of observation proper motion corrected stellar
coordinates were used.

A note on error estimation is in order at this point. As a multi-user
instrument UVES in service mode undergoes frequent setup changes
(central wavelength, slit width, image slicers, etc.). This can
lead to systematic RV shifts of purely instrumental nature
that may not be fully accounted for by our IP modelling approach.
The IP will be more and more affected by setup changes when their
number increases. Conversely, within a given night or on a time scale
of a few nights higher instrumental stability can be expected.
Errors estimated from the scatter in a single night 
are most likely not representative for the long-term error.
On the other hand, errors estimated over time scales of a few nights
may already be affected by real variability of the star. 

%--------------------------------------------------------------
   \begin{figure*}
   \sidecaption
   \includegraphics[width=12cm]{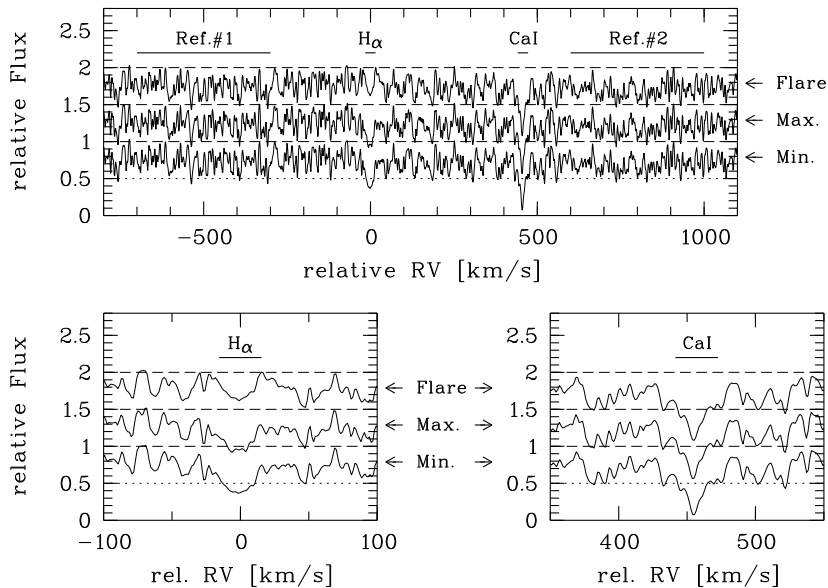}
%
%   \resizebox{\hsize}{!}{\includegraphics{figure3.eps}}
%   \centering
%   \includegraphics[width=10.0cm]{figure3.eps}
%%%   \includegraphics{figure3.eps}
      \caption{~A selection of nightly mean spectra
               of Barnard's star near H$_\alpha ~6563$ and
               CaI~$6573$. Upper panel: three spectra are shown
               vertically offset. They correspond to
               the minimum H$_\alpha $ index (lower spectrum), the maximum
               ``quiescent'' H$_\alpha $ index (middle spectrum offset by
               0.5 flux units), and the flare event (upper spectrum offset
               by 1.0 flux units). Lower panels: Zooms into the upper panel
               on H$_\alpha $ (left) and CaI (right). While a filling-in
               of the H$_\alpha $ line by emission can be seen, the CaI
               line is constant. Horizontal bars indicate the spectral
               regions over which average fluxes are determined for the
               line indices according to Eq.~(5).
              }
%         \label{Fig3}
   \end{figure*}
%
%--------------------------------------------------------------

For this reason we chose to perform all subsequent analysis of our data
on nightly mean values of the DRV
(46 data points) in order to avoid an overestimation of the significances
of possible signals. For the same reason we chose not to use the
propagation of the errors of the individual nightly DRV values to obtain
the error of the nightly mean DRV (which on average for all nights would be 
$1.54~{\rm m~s}^{-1}$); neither did we choose the RMS of the individual nightly
DRV values around the mean for this purpose (which on average would give
$1.39~{\rm m~s}^{-1}$). Instead we chose the {\em mean of the individual
errors} to represent the error of the nightly mean DRV (on average
$2.65~{\rm m~s}^{-1}$). This choice may appear a bit conservative and
arbitrary, but we estimate that it leads to more realistic values for the
nightly errors as it is to a large extent the imperfection of our IP
reconstruction that is reflected in the error of the individual measurement
and should therefore also be reflected in the nightly error.

%---------------------------------------------------------------
   \begin{table}
   \centering
      \caption{
               Models for the UVES data of Barnard's star:
               constant, predicted secular acceleration (S.A.),
               and linear fit. For each model we list the
               slope, the RMS scatter of the data around the model,
               the $\chi ^2$ of the fit, the degrees of freedom (dof),
               and the associated probability.
               The RMS values should be compared with
               the mean internal measurement error of 
               $2.65~{\rm ms}^{-1}$. The last line corresponds to a model
               based on the predicted secular acceleration 
               together with a linear correlation with the 
               ${\rm H}_\alpha $ line strength index defined in Sect.~5.2.
               }
      \vspace{0.2cm}
         \begin{tabular}{llcccl}
            \hline
            \noalign{\smallskip}
Model & Slope & RMS & $\chi ^2$ & dof & $p$ \\
      & $[{\rm ms}^{-1}{\rm yr}^{-1}]$ & $[{\rm ms}^{-1}]$
      &      &     &        \\
            \noalign{\smallskip}
            \hline
            \noalign{\smallskip}
Const.$^{\rm a}$   & $(0.0)$        & 3.93 & 97.2 & 45 & $1.3\cdot 10^{-5}$ \\
S.A.$^{\rm a}$     & $4.499$        & 3.38 & 72.3 & 45 & $0.0065$ \\
Lin.~fit$^{\rm a}$ & $2.97\pm 0.51$ & 3.17 & 63.4 & 44 & $0.029$ \\
Lin.~fit$^{\rm b}$ & $5.15\pm 0.89$ & 2.70 & 29.0 & 28 & $0.41$ \\
             \noalign{\smallskip}
           \hline
            \noalign{\smallskip}
S.A.+${{\rm H}_\alpha }^{\rm c}$ & $(4.499)$ & 2.94 & 54.4 & 43 & $0.12$ \\
            \noalign{\smallskip}
            \hline
         \end{tabular}
\begin{list}{}{}
\item[$^{\rm{a}}$] Data from all 46 nights.
\item[$^{\rm{b}}$] Data from the first 2 years only (first 30 nights).
\item[$^{\rm{c}}$] Data from all nights excluding flare event (45 nights).
\end{list}
   \end{table}
%---------------------------------------------------------------

%------------------------------------------------------------------------------
\section{Results}
%------------------------------------------------------------------------------
\subsection{RV secular acceleration}
%------------------------------------------------------------------------------

Fig.~2 shows the time series of our DRV data of Barnard's star.
Each data point represents the nightly average of the DRV.
The predicted secular acceleration of the RV
(Sect.~3) is overplotted as a
solid line (zero point matched). The dotted line represents a
linear fit to the data.

The best-fit linear trend yields a slope of
$2.97\pm 0.51~{\rm m~s}^{-1}~{\rm yr}^{-1}$ somewhat smaller (by $3.0\sigma $)
than the predicted secular acceleration of 
$4.499~{\rm m~s}^{-1}~{\rm yr}^{-1}$. Clearly, a constant model is rejected
as it would be $5.8\sigma $ away from the fitted slope. For each model
the slope and RMS scatter of the data around it are listed in Table~2
together with the $\chi ^2$, degrees of
freedom, and associated probability.

The relatively small formal probability of the secular acceleration
model $(0.65\% )$ indicates that additional variability is present 
in our data. The linear fit attempts to account for part of this
variability. If only the data of the first 2~years are used for a linear fit
($BJD<2~452~000$, i.e.~the first 30~nights) 
the resulting slope is $5.15\pm 0.98~{\rm m~s}^{-1}~{\rm yr}^{-1}$ which
agrees within $0.95\sigma $ with the predicted secular acceleration
(not shown in Fig.~2).
Around $BJD=2~452~500$ we had intensified the observations in an attempt to
obtain a better sampling of a suspected minimum in the RV residuals from
the secular acceleration model. The presence of this minimum had been suggested
by period analysis and the minimum was clearly found, but it added more weight
to negative residual DRV values thereby reducing the slope of the linear fit.

%------------------------------------------------------------------------------
\subsection{Correlation between H$_\alpha $ line strength and RVs}
%------------------------------------------------------------------------------

In order to find out whether stellar activity is the source of, or a 
contributor to, the observed RV variability we examined the filling-in 
by chromospheric emission of the H$_\alpha $ absorption line in an attempt
to find a correlation with the DRV data. H$_\alpha $ ($6562.808$~{\AA })
is the only activity indicator in the wavelength regime encompassed by
our UVES spectra. Filling-in of the core of this absorption line is an
indicator for chromospheric plage, i.e.~active regions where
magnetic fields are important for the convective properties of the star
and where star spots may also influence the shape of photospheric
absorption lines.

M dwarf spectra are very rich in absorption lines
which blend and do not show a continuum between them which makes it 
difficult to obtain precise absolute line fluxes. Since, however, only
relative fluxes are of interest here we define a suitable line index by

\begin{equation}
   I = \frac{\overline{F_\circ }}
   {0.5*(\overline{F_1} + \overline{F_2})}~,
\end{equation}
where $\overline{F_\circ }$ is the mean spectral flux in a selected
RV interval around the line to be studied. For the ${\rm H}_\alpha $ line
we chose the RV interval $[-15.5,+15.5~{\rm km~s}^{-1}]$ 
centered on the core of the line. $\overline{F_1}$ and $\overline{F_2}$
are the mean fluxes in two reference bandpasses on either side of
H$_\alpha $ (assumed to exhibit only negligible variability with the
appearance of active regions)
for which we chose the intervals $[-700,-300~{\rm km~s}^{-1}]$ and
$[+600,+1000~{\rm km~s}^{-1}]$, respectively. These velocities
are corrected for the barycentric Earth motion (cf.~Sect.~3) and for the
ARV of Barnard's star (Sect.~2).

%--------------------------------------------------------------
   \begin{figure}
   \resizebox{\hsize}{!}{\includegraphics{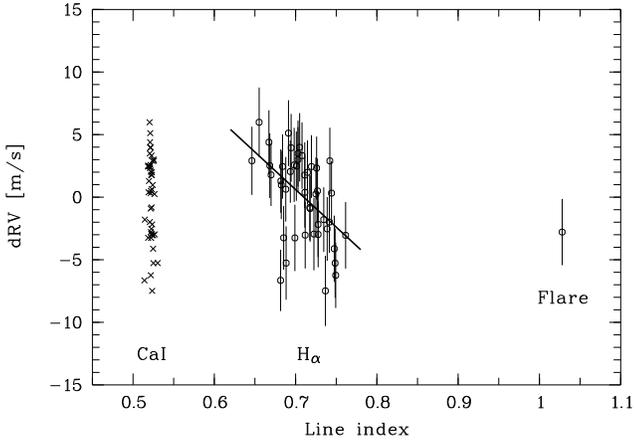}}
%   \centering
%   \includegraphics[width=10.0cm]{figure4.eps}
%%%   \includegraphics{figure4.eps}
      \caption{DRV (after subtraction of the secular acceleration)
        vs.~spectral line index (see text). Circles with RV error
	bars: DRV vs.~${\rm H}_\alpha $ index. Crosses:
	DRV vs.~CaI index. The line represents a linear fit
        to the DRV vs.~${\rm H}_\alpha $ index relation. The rightmost
        data point correpsonds to the flare event.
              }
%         \label{Fig4}
   \end{figure}
%
%--------------------------------------------------------------

For comparison we evaluated the line index for a nearby CaI line at
$6572.795$~{\AA } with low excitation potential that is very strong in 
M dwarfs and whose line strength is not expected to have a significant
dependance on stellar activity. For this line we chose $F_\circ $ over
the RV range $[+441.5,+472.5~{\rm km~s}^{-1}]$
(again relative to ${\rm H}_\alpha $).
Should our data reduction have produced systematic flux offsets,
e.g.~due to imperfect subtraction of the Echelle background, we should see
them in the CaI line.

Fig.~3 shows three of our (nightly mean) spectra of Barnard's star near the
${\rm H}_\alpha ~6563$ and CaI~$6573$ lines. They correspond to the
spectrum with the smallest
${\rm H}_\alpha $ index (labelled ``Min.''), to the spectrum with the
largest ``quiescent'' ${\rm H}_\alpha $ index (labelled ``Max.'') and
to the spectrum of a flare event with an ${\rm H}_\alpha $ index
that is almost $12\sigma $ larger than the mean index of the remainder of the
data (see also Fig.~4). Horizontal bars in Fig.~3 indicate the
spectral regions over which the mean fluxes for Eq.~(5) were determined.

The variation of the ${\rm H}_\alpha $ line strength can be
seen as filling-in by emission, whereas the CaI line is constant.
Interestingly, the smallest ${\rm H}_\alpha $ index was observed
in the night following the flare event. It appears that either the flare 
occured on a quite inactive (visible) hemisphere
or that the flare event has disrupted a major active region
(just as solar flares cause a reconfiguration of the local magnetic
fields). Since we have no data from the night before the flare these 
possibilities must remain speculation.

%--------------------------------------------------------------
   \begin{figure}
   \resizebox{\hsize}{!}{\includegraphics{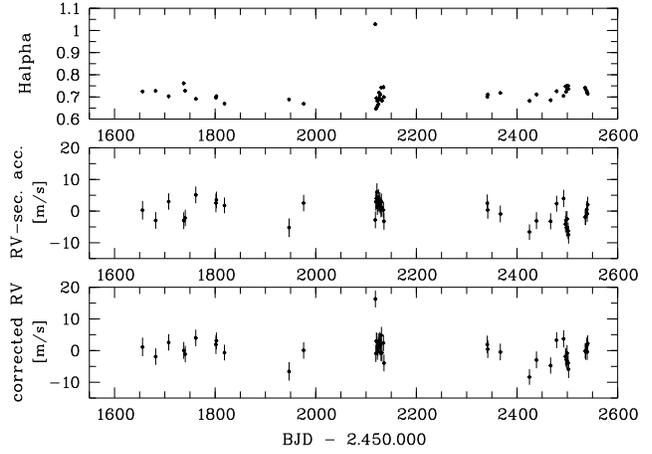}}
%   \centering
%   \includegraphics[width=10.0cm]{figure5.eps}
%%%   \includegraphics{figure5.eps}
      \caption{Comparison of DRV time series and time series of
               ${\rm H}_\alpha $ indices. 
               Top: Time series of ${\rm H}_\alpha $ indices.
               Center: DRV time series (secular acceleration
               subtracted); RMS$=3.39~{\rm m~s}^{-1}$ (excluding
               flare event).
               Bottom: Time series of DRV values corrected for the
               linear correlation with the ${\rm H}_\alpha $ index
               displayed in Fig.~4; RMS$=2.94~{\rm m~s}^{-1}$
               (excluding flare).
              }
%         \label{Fig5}
   \end{figure}
%
%--------------------------------------------------------------

Fig.~4 shows a correlation of our DRV measurements (secular
acceleration subtracted) with the ${\rm H}_\alpha $ and CaI indices.
Just as the DRV values also the line indices were determined in the
individual spectra of each night and then binned into nightly averages.
Again, the CaI line is seen constant 
and independent from the DRV, but the ${\rm H}_\alpha $ line strength is
variable and anti-correlated with the DRV. The flare event is seen only in
${\rm H}_\alpha $, but not in the CaI index. The relative scatter of the
line indices is $0.54\% $ for CaI and $7.6\% $ for ${\rm H}_\alpha $
(or $3.8\% $ when the flare is excluded). The correlation
coefficient is $-0.498$ (flare event excluded). A linear fit to the 
relation of DRV vs.~${\rm H}_\alpha $ index, $I({\rm H}_\alpha )$, is given by
${\rm DRV}=(43\pm 10)~{\rm m~s}^{-1} - 
           (60\pm 15)~{\rm m~s}^{-1}~I({\rm H}_\alpha )$~.
It reduces the total RMS of the DRV data
(flare excluded) from $3.39~{\rm m~s}^{-1}$ to
$2.94~{\rm m~s}^{-1}$. The probability of $\chi ^2$ for this linear fit
is $p=0.12$ rendering this model an acceptable description of the DRV data
(see~Table~2 for all numerical values of the fit).

%--------------------------------------------------------------
   \begin{figure*}
   \sidecaption
   \includegraphics[width=12cm]{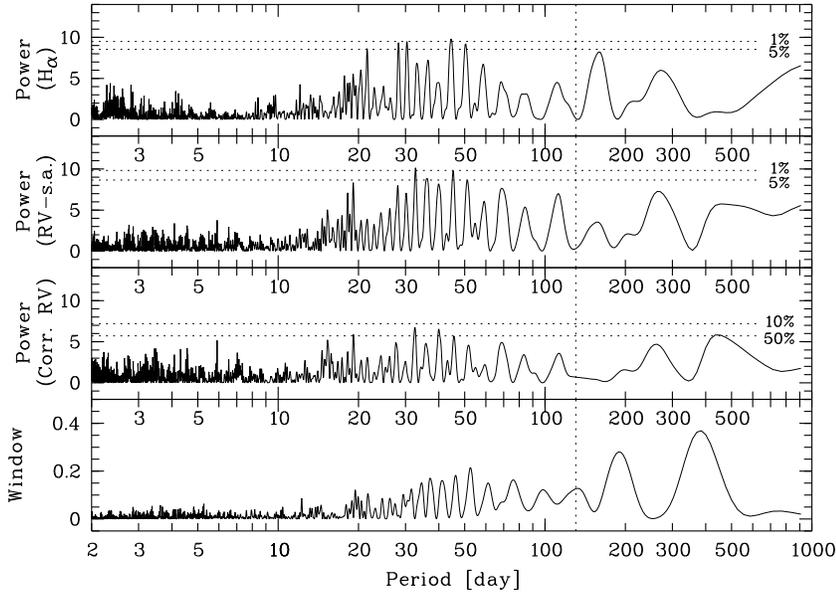}
%
%   \resizebox{\hsize}{!}{\includegraphics{figure6.eps}}
%
%   \centering
%   \includegraphics[width=10.0cm]{figure6.eps}
%%%   \includegraphics{figure6.eps}
      \caption{~Lomb-Scargle periodograms for our data of Barnard's
      star.
      Top panel: ${\rm H}_\alpha $ index.
      Second panel from top: RV data (secular acceleration subtracted).
      Third panel: RV data additionally corrected for the correlation
      with the ${\rm H}_\alpha $ index.
      Bottom panel: window function.
      In each case the flare event was excluded from the data.
      Horizontal dotted lines indicate levels of the FAP ($1\% $ and
      $5\% $ in the upper two panels and $10\% $ and $50\% $ in the
      third). The vertical dotted line markes the supposed
      $130.4~{\rm d}$ rotation period by Benedict et~al.~(1998).
              }
%         \label{Fig6}
   \end{figure*}
%
%--------------------------------------------------------------

A technical note: near ${\rm H}_\alpha $ our spectra still contain
weak iodine absorption lines from the iodine cell (cf.~Sect.~4).
The deepest iodine line in the wavelength regime relevant for our
line indices has a depth of $9.3\% $ relative to the continuum;
about half of this value is more typical for the average line.
% The flux RMS of this iodine line pattern is $1.1\% $.
Due to the barycentric
motion of the Earth, for which the chosen RV intervals are corrected
(with correction extrema of $-26.4~{\rm km~s}^{-1}$ and 
$+26.2~{\rm km~s}^{-1}$) variable parts of the iodine spectrum
contribute to the flux values in Eq.~(5). Calculating the
${\rm H}_\alpha $ and CaI indices for a pure iodine spectrum
(a flatfield exposure taken through the iodine cell) we obtain 
% a peak-to-peak variation of $0.74\% $ and $0.49\% $
an RMS variation of $0.20\% $ and $0.17\% $ for the 
${\rm H}_\alpha $ and CaI index, respectively,
over the range of applied barycentric correction velocities.
This intrinsic uncertainty is much smaller than the observed
scatter of the line indices for our spectra of Barnard's star
($3.8\% $ and $0.54\% $, respectively).
Therefore, our neglection of the contribution of the weak iodine
lines to these spectra does not introduce a major
error in the determination of the line indices.

In Fig.~5 we show a comparison of the time series of our
DRV data (secular acceleration subtracted) with the time series of the
${\rm H}_\alpha $ index. Also included in Fig.~5 is the time series of
the DRV data corrected for the variation with ${\rm H}_\alpha $, 
i.e.~after subtracting the linear fit shown in Fig.~4.

Even if the total scatter of the DRV data is reduced when the
correlation with the ${\rm H}_\alpha $ index is applied for
a correction, the basic variability pattern appears largely unchanged.
Formally, the probability of $\chi ^2$ provides an acceptable model, but
we cannot exclude the possibility that our error estimates were too
conservative (cf.~Sect.~4). We should also note that we do not expect a simple
linear correlation to be sufficient for a complete correction of the
RV data for the effects of stellar activity.

%---------------------------------------------------------------
   \begin{table}
   \centering
      \caption{Period analysis results for the different data sets:
               ${\rm H}_\alpha $ index, RV after subtraction of the secular
               acceleration, RV additionally corrected for the linear
               correlation with the ${\rm H}_\alpha $ index, and
               window function. For the data
               sets that exclude the flare event 
               the period values $P$ of the four highest periodogram peaks
               as well as the pertaining FAP are listed. Only for the
               RV data did we also evaluate period and FAP values with the
               flare event included.
               Similarity of a period value with an integer fraction
               of the photometric period by Benedict et~al.~(1998) is
               indicated in the last column.
               }
      \vspace{0.2cm}
         \begin{tabular}{lrrrrl}
            \hline
            \noalign{\smallskip}
Data set & $P~[{\rm d}]$ & FAP~~ & $P~[{\rm d}]$ & FAP~~ & Comment \\
         & \multicolumn{2}{c}{flare excluded} 
         & \multicolumn{2}{c}{including flare} & \\
            \noalign{\smallskip}
            \hline
            \noalign{\smallskip}
${\rm H}_\alpha $ index &  44.5 & ~0.82\% & & & $\approx P_{\rm ph}/3$ \\
                        &  30.4 & ~1.30\% & & &  \\
                        &  28.2 & ~1.54\% & & &  \\
                        &  50.4 & ~2.06\% & & &  \\
%                        &  21.6 & ~4.49\% & & &  \\
%                        & 159~~ & ~7.96\% & & &  \\
            \noalign{\smallskip}
            \hline
            \noalign{\smallskip}
RV                      &  32.7 & ~0.56\% & 32.7 & 1.56\%  
                        & $\approx P_{\rm ph}/4$ \\
                        &  45.3 & ~1.05\% & 45.3 & 4.27\%  
                        & $\approx P_{\rm ph}/3$ \\
                        &       &         & 19.1 & 5.06\% &  \\
                        &  36.1 & ~4.01\% & 36.0 & 8.26\% &  \\
                        &  51.1 & ~5.49\% &                 &  \\
            \noalign{\smallskip}
            \hline
            \noalign{\smallskip}
RV$-{\rm H}_\alpha $    &  32.5 & 36.81\% & & & $\approx P_{\rm ph}/4$ \\
                        &  40.0 & 45.21\% & & & \\
                        &  19.1 & 75.63\% & & & \\
                        & 440~~ & 77.96\% & & & \\
            \noalign{\smallskip}
            \hline
            \noalign{\smallskip}
Window                  & 380   & ---     & & & $\approx 1~{\rm yr}$ \\
                        & 190   & ---     & & & $\approx 1/2~{\rm yr}$ \\
                        &  52.5 & ---     & & & \\
                        &  46.2 & ---     & & & \\
            \noalign{\smallskip}
            \hline
         \end{tabular}
   \end{table}
%---------------------------------------------------------------

%------------------------------------------------------------------------------
\subsection{Period analysis}
%------------------------------------------------------------------------------

We searched for periodic signals in the ${\rm H}_\alpha$ index, in the
RV residuals from the secular acceleration model, and in the
RV data after additional correction for the linear correlation with the
${\rm H}_\alpha$ index. We used the periodogram by Lomb (1976) and
Scargle (1982) as well as a routine that employs $\chi ^2$ 
fitting of sine waves. Contrary to the Lomb-Scargle
method the sine fitting routine can adapt the RV zero point offset and
also take into account data errors (cf.~Cumming et~al.~1999).
The results of both methods do not differ substantially so that we only
present the results from the Lomb-Scargle periodogram. 
Fig.~6 shows periodograms of the ${\rm H}_\alpha $ index, the RV residuals,
and the RV data corrected for the
correlation with ${\rm H}_\alpha $; the window function is
also shown.

We searched the period range from $2~{\rm d}$ (our minimum temporal resolution)
to $2.42~{\rm yr}$ (our total time baseline). False alarm
probabilities (FAP) for the periodogram peaks were determined with a
bootstrap randomization scheme (e.g.~K\"urster et~al.~1997).
$1\% $ and $5\% $ levels of the FAP are indicated in Fig.~6
for the periodograms of the ${\rm H}_\alpha $ and
RV residuals. After correction
with the ${\rm H}_\alpha $ correlation the power of the RV data was
found considerably reduced; we display $10\% $ and $50\% $ FAP levels
for this data set. 

Table~3 lists the periods and FAPs of the four highest periodogram peaks
for each of the three different data sets excluding the flare event as
well as for the RV data including the flare event (pertaining
periodogram not shown in Fig.~6).
In none of the periodograms is there much power near the supposed
photometric rotation period, $130.4~{\rm d}$, by Benedict
et~al.~(1998) which is also indicated in Fig.~6 as a vertical dotted line.
Nevertheless, the highest peaks tend to coincide with $1/3$ or $1/4$
of the photometric period value which may hint at a common origin.

%--------------------------------------------------------------
   \begin{figure}
   \resizebox{\hsize}{!}{\includegraphics{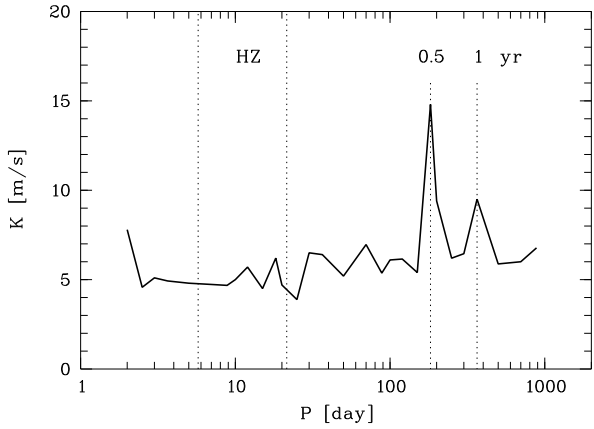}}
%   \centering
%   \includegraphics[width=10.0cm]{figure7.eps}
%%%   \includegraphics{figure7.eps}
      \caption{Upper limits to the RV semi-amplitude $K$ 
               (for circular orbits) as a function
               of orbital period $P$. For each period value these 
               limits correspond to the semi-amplitude at which 
               the artificial signal was found at the
               $FAP\le 1\% $ level for {\em all} 12 test phases.
               Two vertical dotted lines at periods of $5.75$ and 
               $21.5~{\rm d}$ indicate the range of the
               habitable zone around Barnard's star according to
               Kastings et~al.~(1991). Periods of $0.5$ and $1~{\rm yr}$
               are also marked by vertical dotted lines; due to the
               structure of the data sampling relatively high limiting
               amplitudes result at these periods.
              }
%         \label{Fig7}
   \end{figure}
%
%--------------------------------------------------------------

%------------------------------------------------------------------------------
\subsection{Upper limits to companion masses}
%------------------------------------------------------------------------------

Faced with the result that the observed RV variability appears to be 
attributable to stellar activity to a large extent and yields no
evidence for orbiting companions we studied which types of 
companions can be excluded, i.e.~we determined upper
limits to the projected mass $m\sin i$ of orbiting companions
(with $i$ the orbital inclination).
To this end we determined the amplitudes of RV signals
that we would have detected with $99\% $ confidence.

Again bootstrap randomization was employed. In the analysis artificial
RV signals (see next paragraph for details) are added to the observed RV data
which act as a noise term.
% which assumes that
% our RV data are pure noise, i.e.~neglecting any systematic variations.
Since on the one hand
secular acceleration is a well understood effect, but on the other
hand the linear correlation with the ${\rm H}_\alpha $ is probably 
not a complete description of the effect that activity exerts on RV
measurements, we decided to use as the noise term
the secular acceleration subtracted RV data including the 
flare event and uncorrected for the ${\rm H}_\alpha $ correlation.
This data set has a higher rms scatter than the corrected one (see Table~2),
i.e.~it adds more noise to the test signals 
and yields more conservative upper limits.
Note also that the highest peak in the periodogram of this data set
itself (no signal added) appeared at $FAP=1.56\% $ 
(fifth column in Table~3) and would therefore not have passed 
the criterion of $99\% $ confidence that we adopt here.

%--------------------------------------------------------------
   \begin{figure}
   \resizebox{\hsize}{!}{\includegraphics{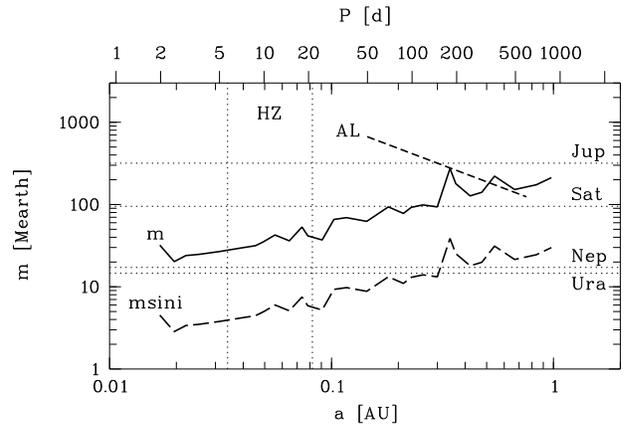}}
%   \centering
%   \includegraphics[width=10.0cm]{figure8.eps}
%%%   \includegraphics{figure7.eps}
      \caption{Upper limits to the mass $m$ of planetary companions 
      in circular orbits around
      Barnard's star plotted vs.~star-planet separation $a$ (bottom)
      and orbital period $P$ (top). 
      Long-dashed line labelled ``msini'': $99\% $ confidence
      limits to the projected companion mass. Solid line
      labelled ``m'': limits to the true companion mass based
      additionally on the
      $99\% $ confidence range for the inclination $i$; i.e.~their combined
      confidence is $98\% $.
%      Dot-dashed line labelled ``NA'': limits to $m\sin i$ from the
%      formula by Nelson \& Angel (1999).
      Short-dashed line labelled ``AL'': astrometric limits to $m$
      from Benedict et~al.~(1999).
      Horizontal thin dotted lines indicate the masses of Jupiter,
      Saturn, Neptune, and Uranus. 
%      According to Kasting et~al.~(1991) the
%      habitable zone lies between the vertical
%      thin dotted lines.
      As in Fig.~7 the habitable zone according to Kasting et~al.~(1991)
      lies between the vertical thin dotted lines.
              }
%         \label{Fig8}
   \end{figure}
%
%--------------------------------------------------------------

Sinusoids were used for the artificial signals
% Artificial data sets were produced by adding sinusoidal signals to our data. 
% Our analysis is therefore strictly valid only for circular orbits.
rendering our analysis strictly valid only for circular orbits.
We used 29 different period values
encompassing the accessible period range $2~{\rm d}-2.42~{\rm yr}$.
At each period 12 equally spaced phases 
were probed, and the amplitude of the sine wave was varied.
The amplitude at which for a given input period
the signal showed up at {\em all} 12 phases at the $FAP\le 1\% $ level
was taken as the minimum detectable amplitude.
We did not require that the
signal appear at the {\em same} period in the power spectrum (due to
spectral leakage it not always did), but that it cause power with a
chance probability $\le 1\% $ {\em somewhere} in the search range.

Fig.~7 shows the resulting limiting RV semi-amplitudes as a function of
orbital period. The median (mean) detectable RV semi-amplitude in the whole
period search range is $5.88~(6.20)~{\rm m~s}^{-1}$; in the habitable
zone it is $4.73~(5.07)~{\rm m~s}^{-1}$. These limits are smaller than
any of the amplitudes found so far for extrasolar planets.

% In HZ:    Mean-K   =  5.0735712 m/s
%           Median-K =  4.730     m/s
%           Min-K    =  4.500     m/s at P =  15.000 d
%           Max-K    =  6.205     m/s at P =  18.262 d
% ----------------------------------------------------
% All data: Mean-K   =  6.1955256 m/s
%	    Median-K =  5.880     m/s
%	    Min-K    =  3.885000  m/s at P =  25.000 d
%	    Max-K    = 14.800     m/s at P = 182.620 d

The amplitude limits were then converted to upper limits on the projected
mass $m\sin i$ of planetary companions in circular orbits. A stellar
mass of $0.16~{\rm M}_\odot $ was adopted from Henry et~al.~(1999).
The limits on $m\sin i$ are shown as a function of period and
star-planet separation in Fig.~8 (long-dashed line).
To obtain limits to the full companion mass $m$ we used the probability
(for random orientation of orbits)
that the orbital inclination $i$ exceeds some angle
$\theta $, given by
$p(i>\theta )=\cos(\theta )$.
In $99\% $ of the cases the true mass does not exceed
the projected mass by more than a factor $7.088$. The solid line in
Fig.~8 shows these mass limits which have a combined
probability (that of signal detection and that of the inclination range)
of $0.99^2\approx 98\% $.
The short-dashed line labelled ``AL'' shows astrometric mass limits
by Benedict et~al.(1999) which are $99\% $ confidence limits, but
allow for a $5\% $ miss rate for astrometric signals.

As seen in Fig.~8 the limits on $m\sin i$ increase roughly from a few
Earth masses at the smallest separations to 
$\approx 30~{\rm M}_{\rm Earth}$ at larger separations.
At periods of $1/2~{\rm yr}$ and $1~{\rm yr}$ we obtain somewhat higher limits
because of lower detection capabilities due to the data sampling window.

%-----------------------------------------------------------------------------
\section{Discussion}
%-----------------------------------------------------------------------------
\subsection{RV secular acceleration}
%-----------------------------------------------------------------------------

We clearly find a positive trend in our DRV data of Barnard's star
and interpret it as mainly due to the secular acceleration of the RV. 
The formal discrepancy between a linear fit and the predicted
secular acceleration value can be attributed to stellar acticity,
in particular to incomplete sampling of the time scale(s) on which
the activity varies.

To our knowledge this is the first reported observation of stellar
RV secular acceleration. With time we expect the observed trend to
conicide more with the predicted acceleration as activity effects
cancel out more.

%-----------------------------------------------------------------------------
\subsection{The activity-related RV  signal: Evidence for convective redshift?}
%-----------------------------------------------------------------------------

Active regions, which in our data reveal their appearance through the
filling-in of the ${\rm H}_\alpha $ line core by emission,
can affect the apparent
stellar RV as measured in lines of photospheric origin by the effects of
star spots or chromospheric plage regions. Compared to the inactive
photosphere both spots and plage
produce a local change of brightness and spectrum and also inhibit
stellar convection locally by magnetic fields.

The primary effect of dark star spots at different locations on the
visible stellar disk is a reduction of the observable flux at the
instantaneous RV of the spot(s), as spot temperatures $T_{\rm s}$
are considerably smaller than photospheric effective temperatures
$T_{\rm eff}$ and $(T_{\rm s}/T_{\rm eff})^4 \ll 1$. Therefore, the effects
that spots have a different spectrum and that convection is suppressed
inside the spot have small contributions to the integrated stellar spectrum
compared to the general flux reduction in a small RV regime.
If there is sufficient rotational broadening of the absorption lines,
then this will lead to
a change in the center-of-gravity of the line thus mimicking an RV change.
The effect is more pronounced when spots are displaced from the limb of the
visible stellar disk where their instantaneous RV (relative to the RV
of the star as a whole) is high. The effect disappears when spots are
on the sub-observer meridian of the star where the relative RV is
zero (or when they are on the invisible hemisphere of the star). 

In a model that assumes that photospheric spots are
surrounded by chromospheric plage the following picture emerges.
Minimum RV shift, i.e.~most active regions hidden from view or
most active regions visible and concentrated towards the sub-observer
meridian, should be accompanied by either small or large amounts of the
filling-on of the ${\rm H}_\alpha $ line core by emission. Maximum
RV shift, i.e.~active regions concentrated on either side of the
sub-observer meridian, should be seen when intermediate to large
amounts of ${\rm H}_\alpha $ filling-in are observed 
(see also Paulson et al.~2002; Saar \& Fischer 2000).

Consider the simple case of a single small
spot on an otherwise featureless rotating star.
Let $A_{\rm s}$ be the spot area and let
$\phi$ be the rotation angle of the spot from the sub-observer
meridian. The RV perturbation $\Delta v_{\rm r}$ then scales as
$\Delta v_{\rm r}(\phi)
\propto -A_{\rm s} f(\phi ) g(\phi ) v \sin i$, where 
$A_{\rm s}$ is the spot area, $f(\phi )$ is a function describing 
the visibility and foreshortening due to projection of the spot,
hence $A_{\rm s} f(\phi )$ is the projected spot area, $g(\phi )$ is a
function describing the deflection of the stellar RV caused by the
spot as a function of stellar rotation phase $\phi $, and $v \sin i$
is the projected rotational velocity of the star. The minus sign
indicates that the RV shift of the line as a whole is
opposite to the instantaneous RV of the dark spot due to the flux
reduction it produces.

Both $f(\phi )$ and $g(\phi )$ are periodic functions of $\phi $ which
possess special symmetries around $\phi =0$, 
i.e.~the phase where the spot is on the sub-observer meridian.
The projection function $f(\phi )$ is mirror symmetric with respect to
this phase (similar to a cosine function)
while the RV deflection $g(\phi )$ has a rotational
symmetry around this phase (just like a sine function does). In
addition $g(\phi )$ has zero mean.
\footnote{For the RV deflection to be a fully rotationally symmetric
function with zero mean
the absorption line(s) under study must have symmetric
profiles which is a good approximation.} 
Due to these properties of $f(\phi )$ and $g(\phi )$
the RV perturbation averaged over phase $\phi $ vanishes, 
i.e.~$\langle \Delta v_{\rm r}\rangle =0$.
On time scales longer than the stellar rotation period the mean spot
related RV perturbation is therefore uncorrelated with the visibility
of spots and spot coverage as indirectly evidenced by ${\rm H}_\alpha $ 
emission from the surrounding plage. It is only the scatter in the
instantaneous RV deflection which correlates with spot visibility and
area, but not the mean value.

In a plot of DRV vs.~${\rm H}_\alpha $ index one should thus expect
for a single active region (small spot plus surrounding plage) 
a closed curve with a shape that resembles a ``reclining raindrop'';
\footnote{The shape of a ``reclining raindrop'' is obtained when the
popular picture (not the physical one) of a raindrop is rotated
counterclockwise by $90^\circ $.}
if several active regions are present the plot contains 
a superposition of curves of this shape. If plage regions exist
without underlying spots, the ${\rm H}_\alpha $ index will be further
enhanced when they are visible without any spot-related effect on the RV,
thus blurring the plot along its horizontal axis.
In any case, if spots are responsible for the RV
variability the correlation coefficient of the RV vs.~${\rm H}_\alpha $
relation should be close to zero.
This is not evident in Fig.~4 so that star spots cannot be the
dominant contributor to the observed correlation.

Estimates of the maximum spot contribution to $\Delta v_{\rm r}$
are consistent with this. Saar \& Donahue (1998) show that
the spot-induced RV amplitude (in ${\rm m~s}^{-1}$) is
$K_{\rm s} \approx 6.5 f_{\rm s}^{0.9} v \sin i$, where $f_{\rm s}$ 
is the differential spot filling factor (in \%) and $v \sin i$ is in 
${\rm km~s}^{-1}$ (see also Hatzes 2002). Since the stellar radius
is roughly $R \approx 0.2~{\rm R}_\odot $ and the rotation period is
$P_{\rm rot} \ge 32.5~{\rm d}$
(assuming that the period of highest power in the periodogram of the
RV data in Table~2
is either the rotation period or a fraction thereof), one obtains
$v \sin i \le 0.31~{\rm km~s}^{-1}$.  
If the spot-related brightness variation of the star is still similar
to that found by Benedict et~al.~(1998), 
i.e.~$\Delta V \approx 0.01$, implying a change in spot filling factor of
$f_{\rm s} \approx 0.92\% $, we get 
$K_{\rm s} \leq 1.9~{\rm m~s}^{-1}$, and the associated spot-induced scatter
$\sigma _{\rm s}\approx K/\sqrt 2\le 1.3~{\rm m~s}^{-1}$,
far lower than the variation seen here.
\footnote{Subtracting in quadrature the total scatter of the RV data after
subtraction of the secular acceleration, $3.38~{\rm m~s}^{-1}$, and the
average internal error of ${2.65}~{\rm m~s}^{-1}$ (conservative
estimate; cf.~Sect.~4) we obtain an excess scatter of
$2.10~{\rm m~s}^{-1}$.}
This is further indication, under the assumptions we have made, that
spots are not the main contributor for the RV variability in Barnard's star.

The effects of bright chromospheric plage regions on the integrated
spectrum of the star are quite different from those of spots.
In the visual the plage-photosphere brightness contrast is small and
does not exert a strong effect on photospheric absorption line shapes.
Also such an effect would cause similar cyclic RV shifts as spots
(with an inverted sign) and would not lead to a correlation with the RV.
However, the brightness of plage lets them contribute their own spectrum
to the overall stellar spectrum. For the same reason the suppression
of convective flows in the plage magnetic fields can lead to
observable effects.

In solar-type stars one usually observes the plage spectrum in the
form of enhanced emission in those lines in which chromospheric
emission is most readily seen, such as CaII~H+K or the Balmer lines.
This emission is usually blueshifted with respect to the photospheric
line component and its strength increases with activity level thereby 
changing the degree of line asymmetry.
If a similar effect also ocurrs in other photospheric lines (perhaps to
a smaller degree), then it could affect the measured RV.
So far this has not yet been conclusively demonstrated.

However, in our ${\rm H}_\alpha $ line profiles of Barnard's star we
find no evidence for a distinct behaviour of the blue or red parts of
the line. To this end we measured
the ratio of the mean fluxes in the RV intervals
$[-35,0~{\rm km~s}^{-1}]$ and $[0,+35~{\rm km~s}^{-1}]$
around ${\rm H}_\alpha $ (cf.~Fig.~3) and found it
constant within $1.9\% $ rms and uncorrelated
with both our RV data and the ${\rm H}_\alpha $ index. If a variable 
differential emission enhancement between the
blue and red parts of the ${\rm H}_\alpha $ line of Barnard's star
exists, then its degree is below our measurement precision.

Clearly, we cannot draw direct conclusions from this finding on the
contribution of the chromospheric plage spectrum to the absorption
lines relevant for our RV measurements. Nevertheless a tentative assumption
that such a contribution is small does not seem unreasonable.

Then what about the suppression of the local convective velocity
pattern by magnetic plage? 
% The second effect that influences the measured RV, suppression of the local
% convective velocity pattern by magnetic plage, proves to be more
% important for the measured RV. 
% Since plage
% are bright and cover larger areas than spots ($A_{\rm p}\gg A_{\rm s}$)
% this effect should predominantly occur in plage regions for which the
% ${\rm H}_\alpha $ index is a direct indicator. 
Local suppression of
convection alters the center-of-gravity of photospheric absorption lines
(see Cavallini et~al.~1985 for the solar case).
If the fundamental pattern is
convective blueshift, as seen in main sequence stars of types G or K, then any
local suppression of the pattern produces a net redshift of the lines
when active regions appear.
\footnote{For reasons of the conservation of the energy flux suppressed
convection in magnetically active regions should in principle be
compensated by an increase outside of them. To our knowledge searches
for ``bright rings'' around sunspots, where the ``missing energy'' was
expected to emerge, have not been successful. In analogy with the Sun
% As concerns the net shift effect in photospheric absorption lines 
we implicitely assume that the reduced convective line shift due to
the local suppression of convection in plage dominates over the effect
of possibly increased convection outside. 
% For a review of line asymmetries
% and wavelength shifts due to stellar convection see Dravins (1999) and
% references therein.
}

To repeat our basic assumptions, 
we consider the case in which that part of the spectrum that is relevant for
our RV measurements is similar when emitted either from plage or from the
inactive photosphere, both in continuum brightness and absorption line
strengths and shapes. 
% We assume that the part of the spectrum relevant for our RV measurements
% is similar when emitted from plage or from the inactive photosphere, both in
% continuum brightness and absorption line strength. 
Then we find that
for each individual plage region the plage induced RV perturbation scales as 
% $\Delta v_{\rm r}(\phi) \propto -A_{\rm p} f(\phi ) [v_{\rm cs}(\phi)
% +g(\phi) v \sin i]$, where $v_{\rm cs}(\phi)$ is the convective shift.
$\Delta v_{\rm r}(\phi) \propto  -A_{\rm p} v_{\rm cs}(\phi ) f(\phi )$,
where $A_{\rm p}$ is the surface area of the plage, 
$v_{\rm cs}(\phi )$ is the convective blueshift in the absence of plage,
and $f(\phi )$ describes the visibility and foreshortening of the
plage area. Averaging over the rotation phase $\phi $ we
obtain $\langle \Delta v_{\rm r} \rangle 
\propto -A_{\rm p} \langle v_{\rm cs}(\phi ) f(\phi )\rangle $.
Since neither $v_{\rm cs}(\phi )$ nor $f(\phi)$ have zero mean the
plage induced RV redshift does not vanish and increases with plage
area. If more than one plage region is visible, their contributions
are additive.

In this case there should be a clearer correlation between the RV and
the total plage coverage as observable in the ${\rm H}_\alpha $ index.
This seems to be similar to what we see in Fig.~4, however, with
opposite sign. Our figure reveals a net blueshift when the coverage of the
star with plage regions increases which can be explained, if the
convection properties in Barnard's star produce convective redshift as
the fundamental pattern, i.e.~in regions not dominated by magnetic fields.
% \footnote{It remains to be seen whether this property can also be found
% in other M dwarfs.}
If this fundamental pattern is suppressed,
a net blueshift will be the result, as observed.
%pretty much as we see it.

The possibility that M dwarfs have convective redshifts may extend a trend
(by way of extrapolation) seen in recent work.
Gray (1982) found that bisectors
in dwarfs become less curved (``C shaped'') and more vertical
(see also Saar \& Bruning 1990), and also less blueshifted with 
decreasing effective temperature $T_{\rm eff}$. Although Gray's velocity
zero-point was somewhat uncertain, his data possibly
hints that by mid-K main sequence stars, the net convective blueshift
may actually change sign to a convective redshift (see also Dravins 1999).

If convective redshift prevails in Barnard's star, then the majority of the 
photospheric absorption lines relevant for our RV measurements seems to
form in regions of convective overshoot. Another
possibility is that the details of the convection pattern, i.e.~the
interplay of the involved flow velocities, contrasts and surface areas of
granules (where material rises) and inter-granular lanes (where the
gas flows down) cause a net redshift of the relevant lines. 
% Detailed 3-D hydrodynamic models are needed to clarify this.

Theory also supports the idea of a decreasing convective blueshift with
decreasing $T_{\rm eff}$ (Dravins \& Nordlund 1990, Asplund et~al.~2000).
Although analogous line profile calculations for 3-D
hydrodynamic M dwarf models have not yet been published, the models
themselves (Ludwig et~al.~2002) show low contrast 
between upflowing and downflowing regions,
$C({\rm dM})\approx 1.1\% $, and low RMS vertical velocities of up to 
$\sigma_\bot({\rm dM}) \approx 0.24~{\rm km~s}^{-1}$.
Analogous solar models (see also Ludwig et~al.~2002)
yield contrasts $C(\odot)\approx$16\% and RMS vertical velocities
$\sigma_\bot(\odot)\approx 2.6~{\rm km~s}^{-1}$. 
Adopting these values permits us to make a rough estimate of the expected
RV shift due to convective motions, $v_{\rm cs}$, if we assume that it 
scales with $\sigma_\bot$. Then we can expect (including the sign change)
\begin{equation} 
v_{\rm cs}({\rm dM}) = -\frac{\sigma_\bot({\rm dM})}{\sigma_\bot(\odot)}~
  \frac{(1-C({\rm dM}))}{(1-C(\odot))}~v_{{\rm cs}}(\odot)~.
\end{equation}
% $v_{\rm cs}({\rm dM})\approx -0.1 v_{\rm cs}(\odot)$. 
With $v_{\rm cs}(\odot) \approx -300~{\rm m~s}^{-1}$ (e.g.~Dravins 1999), 
this implies a convective redshift of 
$v_{\rm cs}({\rm dM})\approx 33~{\rm m~s}^{-1}$ in the absence of plage.
If plage inhibit the convection at variable amounts as they
appear, grow, shrink, disappear, or rotate out of view,
the possible variation of the convective shift cannot
exceed $\Delta v_{\rm cs}({\rm dM})\approx 33~{\rm m~s}^{-1}$.

This value is consistent with our observed (smaller) range of RV variation of
$6.9\pm 1.7~{\rm m~s}^{-1}$ as obtained from the linear fit to the
relation between DRV and ${\rm H}_\alpha$ index (Fig.~4). We thus estimate
that the variation in the surface filling factor of visible plage is 
approximately $(21\pm 5)\% $.

% which is then 
% the maximum convective shift possible.
% $\Delta v_{\rm r}$ possible, i.e., $A_{\rm p} = 1$), 
% consistent with the observed value of
% $\Delta v_{\rm r} \approx 5$ m s$^{-1}$ (and further suggests that 
% $\Delta A_{\rm p}\approx 0.17$).

% Both observations and theory therefore imply that magnetic plage, which
% inhibits convective motions, should locally reduce the convective
% blueshifts by progressively smaller absolute amounts as $T_{\rm eff}$
% decreases, down to $T_{\rm eff} \approx 5000$ K. Below this $T_{\rm eff}$,
% since it is possible that convective {\em redshifts} are
% typical, plage would presumably reduce {\em them}. Thus the presence
% of identical amounts of plage will induce progressively larger net
% effective redshifts (relative to the plage-free case) for 
% $T_{\rm eff}>5000~{\rm K}$, and net effective blueshifts for
% $T_{\rm eff} < 5000{\rm K}$ (if
% the assumption of convective redshifts in this $T_{\rm eff}$ regime is
% valid). The latter scenario can then explain Fig.~4. On shorter
% timescales, rotational ``reclining raindrops" can be expected,
% explaining much of the scatter, and perhaps some of the arc-shaped
% excursions around the fit line in Fig.~4.  Long-term, these summed
% ``raindrops" (one per active region) can be seen shifting in
% $v_{\rm r}$ and activity as the mean activity level changes.  It is these
% long timescale activity level changes which drive the correlation seen in 
% Fig.~4.

%-----------------------------------------------------------------------------
\subsection{Planetary companions}
%-----------------------------------------------------------------------------

Our detection capabilities (Figs.~7 and 8) with UVES are currently leading
among the ongoing precision RV surveys.
To our knowledge the mass upper limits determined in Sect.~5.4 for
planetary companions in circular orbits are the lowest published so far.

For circular orbits we find no planets with
$m\sin i> 0.12~{\rm M}_{\rm Jupiter}$ and
$m> 0.86~{\rm M}_{\rm Jupiter}$ over the range of separations 
accessible to our RV data ($0.017-0.98~{\rm AU}$)
corresponding to periods of $2-885~{\rm d}$.
In the habitable zone around Barnard's star
(separations of $0.034-0.082~{\rm AU}$ and periods 
of $5.75-21.5~{\rm d}$), we can even exclude
planets with $m\sin i> 7.5~{\rm M}_{\rm Earth}$
and $m> 3.1~{\rm M}_{\rm Neptune}$.

We cannot fully exclude the possibility that we may have missed planets 
with higher masses in high-eccentricity orbits. Note, however, the
astrometric limit in Fig.~8 (from Benedict et~al.~1999).
Taking into account eccentric orbits in our simulations would make the
calculations prohibitively expensive, because of two additional free 
parameters (eccentricity and longitude of periastron). For further 
discussion see Endl et~al.~(2002).

%-----------------------------------------------------------------------------
\section{Conclusions}
%-----------------------------------------------------------------------------
%
\begin{enumerate}
\item We have clearly discovered a linear trend in the RVs of
  Barnard's star which we attribute to the 
  RV secular acceleration of the star.
\item The residuals from the secular acceleration are correlated
  with stellar activity to a large extent.
\item From the sign of this correlation we propose that the
  fundamental (magnetic field free) convection pattern in this M dwarf
  is convective redshift as opposed to the blueshift seen in earlier
  spectral type stars.
\item A rough calculation based on the convective redshift assumption
  leads us to estimate that the variation of the
  visible plage coverage is about $20\% $. 
\item From our RV data we infer the so far lowest published mass 
  upper limits for planets in cicrular orbits around their host star.
  We exclude planets with
  $m\sin i> 0.12~{\rm M}_{\rm Jupiter}$ and
  $m> 0.86~{\rm M}_{\rm Jupiter}$ over separations of
  $0.017-0.98~{\rm AU}$.
  In the habitable zone around Barnard's star,
  i.e.~$0.034-0.082~{\rm AU}$, our RV data exclude
  planets with $m\sin i> 7.5~{\rm M}_{\rm Earth}$
  and $m> 3.1~{\rm M}_{\rm Neptune}$.
\end{enumerate}

%------------------------------------------------------------------------------
\begin{acknowledgements}
A large part of this work was done while MK and FR were at
ESO, Chile, whose support is gratefully acknowledged.
We thank the ESO OPC and the ESO DDTC for generous allocation of
observing time.
The help of the Science Operations Team at Paranal Observatory was
very important. We thank D.~Dravins for valuable discussions on the
effects of plage on RVs.
ME and WDC achnowledge support by NASA grant NAG5-9227 and
NSF grant AST-9808980. SE is supported under Marie Curie Fellowship
contract no.~HDPMD-CT-2000-5. SHS is supported by NASA Origins grant 
NAG5-10630.
\end{acknowledgements}

%------------------------------------------------------------------------------

%
\end{document}